# PROSPECTS OF APPLICATION OF SEMI-DEFINITE PROGRAMMING TO DETERMINE ORBITAL PARAMETERS OF THE BINARY SYSTEMS OBSERVED AT BOSSCHA OBSERVATORY

Budi Dermawan, Moch. Irfan, Suryadi Siregar, Denny Mandey, Hanindyo Kuncarayakti, Djoko Suprijanto

*Institut Teknologi Bandung, Indonesia*

***Abstract.*** *Most methods of orbit determination are often difficult for numerical implementations since they are developed before the computer era. The recently developed mathematical technique of semi-definite programming (SDP) has been implemented for many problems in scientific fields including astrometry. This is a good opportunity to resolve orbits of binary systems located in the southern hemisphere since more than seventy years Bosscha Observatory had been continuously conducting observations of binary systems. Here we describe prospects of application of SDP for deriving orbital parameters of binary systems using data supplied by Bosscha Observatory that has been published in the Centre de Donnees astronomiques de Strasbourg. This study will support observers at Bosscha Observatory to appropriately select target stars belong to binary systems for their ongoing researches. Since SDP is a powerful scheme, free trial-and-error and human-independent judgment, we suggest that SDP may become a standard method for determining orbital parameters of binary systems.*

## 1 Introduction

Numerous literatures have been published regarding on how to resolve orbital parameters of visual binaries. Although visual binaries have been studied for more than 200 years, it is still possible to improve the method because most of the methods are severely dated. They were developed before the computer era.

More than a century ago a well-known method to determine orbital parameters of binaries was coined by Thiele and improved in the last several tens years by Innes and van den Bos [12]. Some other methods deserve to be briefly cited here: the Fourier expansion [8], the Docobo method [4], the adoptive-grid search [5], the simulated annealing [10], the simplex algorithm [7], the Koval'skij-Olević method [9], and recently the inversion formula or Asada method [1]. Two major minors of some old methods are the facts that they need a kind of trial-and-error procedure and a subjective judgment of the ellipse fitting problem.

A mathematical technique for optimization, namely Semi-Definite Programming (SDP), pioneered by [11] has made a breakthrough. Advantages offered by this method are being analytical, solving for a unique ellipse, involving the areal velocity law, being robust, and yielding mean errors of the quantities [2].

Regarding the ellipse fitting problem to binary data, no guarantee that the conic section calculated will be an ellipse, both for homogeneous and non-homogeneous forms. Known procedures suffer from problems of divergences. The goodness-of-fit criterion was improved by [3] in order to obtain a unique ellipse and robust.

## 2 SDP Procedure

According to [11] SDP can be written as

$$C * X = \min, \text{ subject to } A_k * X = b_k \quad (k = 1, \cdots, p)$$

where $p$ denotes number of conditions or unknowns. $C$, $X$, and $A_k$ are symmetric $p \times p$ matrices, $b_k$ are scalars, $X$ is semi-definite (real eigenvalues and non-negative), and the asterisk follows $G*H = \text{trace}(G,H)$.

Relative to the primary star, orbit of the secondary star is an ellipse $ax^2 + by^2 + cxy + dx + fy + l = 0$, where $x$ and $y$ are rectangular coordinate converted from data of the separation $\rho$ and the position angle $\theta$ of a binary system. In more symmetric notation

$$(x \ y)\begin{pmatrix} a & c/2 \\ c/2 & b \end{pmatrix}\begin{pmatrix} x \\ y \end{pmatrix} + (d \ f)\begin{pmatrix} x \\ y \end{pmatrix} + l = 0.$$

Following [2] define vectors $v_i = (x_i, y_i)^T, w = (h, k)^T$ then the matrix equation above becomes

$$(v_i - w)^T A(v_i - w) + ah^2 + bk^2 + dh + fk + l = 0.$$

Let $u$ be a column vector of ones of sizes $m$ and $\gamma$ a column $m$-vector of positive scalars. Define

$$g_i = \begin{bmatrix} A & \begin{pmatrix} h/2 \\ k/2 \end{pmatrix} \\ (h/2 \ k/2) & \gamma_{1,i} \end{bmatrix}$$

Then the ellipse fitting problem becomes $u^T \gamma_1 = \min$, subject to

$$\gamma_{1,i} - g_i > 0, i = 1, \cdots, m,$$
$$\gamma_{1,i} + g_i > 0, i = 1, \cdots, m,$$
$$\gamma_{1,i} > 0, i = 1, \cdots, m,$$
$$A \succ 0, \ \text{trace}(A) = 1.$$

The constant (C) areal velocity law states $\rho^2 \dot{\theta} = C$. C is related to the mean daily motion (the orbital period can then be calculated) and the projected semi-major and –minor axes which the original ones can be derived. The improved goodness-of-fit criterion [3] now must be modified to include the areal velocity constant.

The SDP problem of a binary system becomes $u^T \begin{pmatrix} \gamma_1 \\ \gamma_2 \end{pmatrix} = \min$, subject to

$$\gamma_{1,i} - g_i > 0, i = 1, \cdots, m, \quad \gamma_{2,i} - g_i > 0, i = 1, \cdots, m,$$
$$\gamma_{1,i} + g_i > 0, i = 1, \cdots, m, \quad \gamma_{2,i} + g_i > 0, i = 1, \cdots, m,$$
$$\gamma_{1,i} > 0, i = 1, \cdots, m, \quad \gamma_{2,i} > 0, i = 1, \cdots, m,$$
$$A \succ 0, \ \text{trace}(A) = 1.$$

## 3 Application of SDP [2]

There is a good example to demonstrate the powerful of SDP scheme [2]. The binary system is 24 Aquarii whose collected data is unusual and "recalcitrant". The following figures are adapted from [2] to illustrate the application of SDP to the rather disorder data.

Figure 1 demonstrates a robustness of the ellipse fitting result (crosses) with the highly discordant of data points a, b, and c.

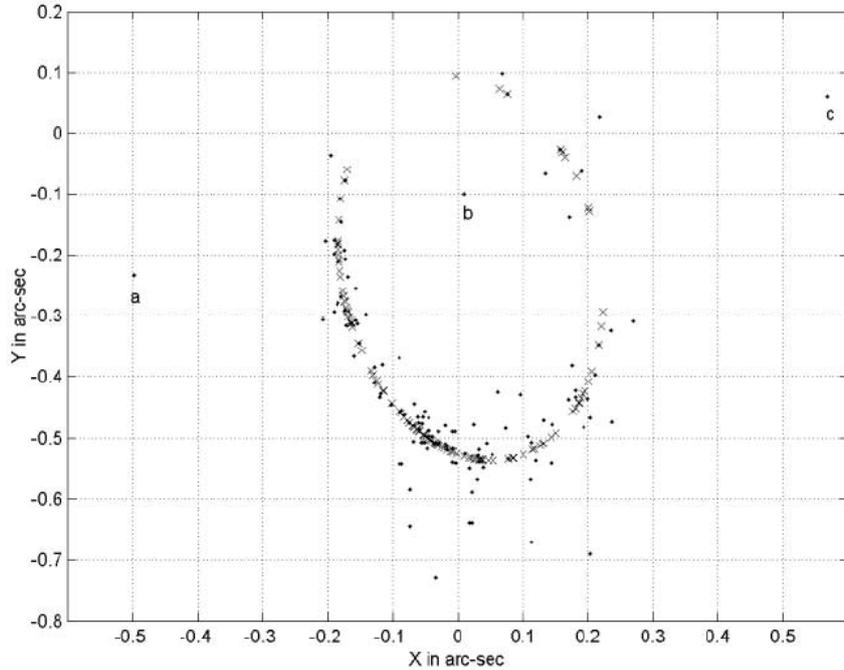

**Figure 1.** Observed data (dots) and calculated ellipse (crosses) of a binary system 24 Aquarii. This figure is adapted from [2].

The need to incorporate statistical weights can also be accommodated by SDP. To do this we need to change the vector $u^T$ from a $2m$ vector of ones to a $2m$ vector of the weights, repeating twice the weights [2].

Figure 2 in the following demonstrates how SDP can derive an ellipse from limited data points taken from the speckle observations, including a discordant point d.

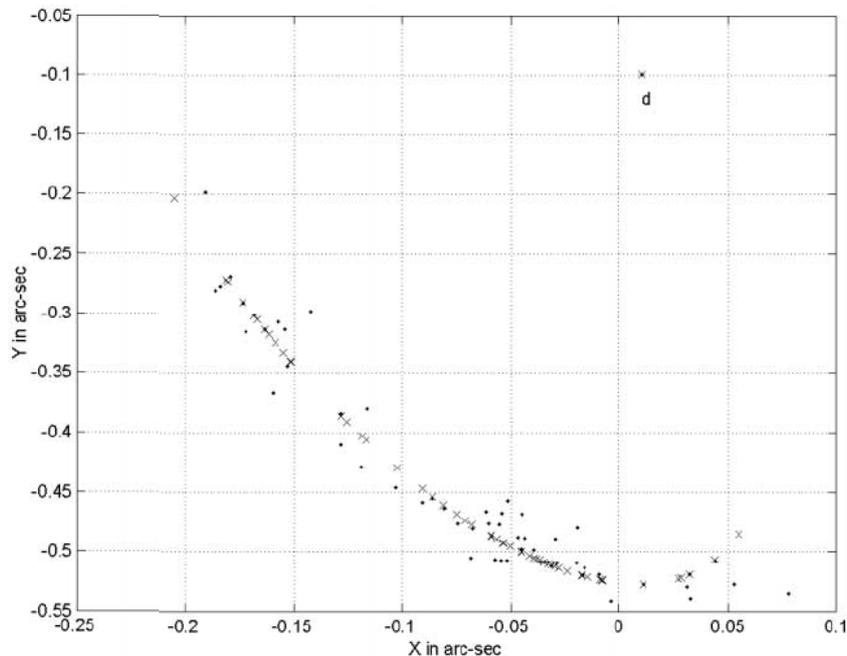

**Figure 2.** Orbit derived from limited data taken from speckle observations. This figure is adapted from [2].

SDP demonstrates a powerful scheme without human intervention of judgment and free trial-and-error step. Discordant data do not alter an ellipse obtained by SDP. SDP is useful for deriving orbital parameters for obtaining the preliminary orbit of a binary system whose number of data is small.

It is now a good opportunity to revisit data of binary systems collected from Bosscha Observatory [6]. Since the data is about 600 systems, systematic procedures should be developed in order to optimally gain the results.

## 4 Summary

SDP proves a flexible mathematical tool to study binary systems because it permits one to mix the norms for the reductions of the various classes of data, calculates the global minimum of the reduction criterion, and presents no problem with convergence to the minimum [2].

With such proofs we suggest that SDP is a powerful tool and more reliable. It may become a standard tool for resolving orbital parameters of binary systems.

## Acknowledgment

We would like to thank to the Organizing Committee and the Astronomy Study Program to make possible the authors to participate and for their supports in this event.

BUDI DERMAWAN, SURYADI SIREGAR
Astronomy Research Division & Bosscha Observatory,
Faculty of Mathematics and Natural Sciences,
Institut Teknologi Bandung, Indonesia
E-mail: budider@as.itb.ac.id, suryadi@as.itb.ac.id

MOCH. IRFAN, DENNY MANDEY, HANINDYO KUNCARAYAKTI
Bosscha Observatory,
Faculty of Mathematics and Natural Sciences,
Institut Teknologi Bandung, Indonesia
E-mail: m.irfan@as.itb.ac.id, eps_eridani_9@yahoo.co.uk, kuncarayakti@gmail.com

DJOKO SUPRIJANTO
Combinatorial Mathematics Research Division,
Faculty of Mathematics and Natural Sciences,
Institut Teknologi Bandung, Indonesia
E-mail: djoko@math.itb.ac.id